\documentclass[sigconf]{acmart}
\renewcommand\footnotetextcopyrightpermission[1]{} 
\usepackage{booktabs} 
\usepackage{algorithm}
\usepackage[noend]{algpseudocode}
\usepackage{caption}
\usepackage[margin=20pt]{subcaption}
\usepackage{hyperref}
\usepackage{enumerate}
\usepackage{mymacros}
\usepackage{balance} 
\usepackage{paralist}
\usepackage[font={small,it},skip=0.5pt]{caption}
\usepackage{setspace}

\acmConference{ExP}{Microsoft}{2018} 
\begin{document}
\title{Applying the Delta Method in Metric Analytics: \\ A Practical Guide with Novel Ideas}

\author{Alex Deng}
\affiliation{
  \institution{Microsoft Corporation}
  \city{Redmond} 
  \state{WA} 
  \postcode{98052}
}
\email{alexdeng@microsoft.com}

\author{Ulf Knoblich}
\affiliation{%
  \institution{Microsoft Corporation}
  \city{Redmond} 
  \state{WA} 
  \postcode{98052}
}
\email{ulfk@microsoft.com}

\author{Jiannan Lu}
\authornote{The authors are listed in alphabetical order.}
\affiliation{
  \institution{Microsoft Corporation}
  \city{Redmond} 
  \state{WA} 
  \postcode{98052}
}
\email{jiannl@microsoft.com}

\begin{abstract}

During the last decade, the information technology industry has adopted a data-driven culture, relying on online metrics to measure and monitor business performance. Under the setting of big data, the majority of such metrics approximately follow normal distributions, opening up potential opportunities to model them directly without extra model assumptions and solve big data problems via closed-form formulas using distributed algorithms at a fraction of the cost of simulation-based procedures like bootstrap. However, certain attributes of the metrics, such as their corresponding data generating processes and aggregation levels, pose numerous challenges for constructing trustworthy estimation and inference procedures. Motivated by four real-life examples in metric development and analytics for large-scale A/B testing, we provide a practical guide to applying the Delta method, one of the most important tools from the classic statistics literature, to address the aforementioned challenges. We emphasize the central role of the Delta method in metric analytics by highlighting both its classic and novel applications. 

\end{abstract}


\maketitle

\section{Introduction}\label{sec:intro}

\subsection{Background}

The era of big data brings both blessings and curses \citep{fan2014challenges}. On one hand, it allows us to measure more accurately and efficiently, to study smaller and more subtle effects, and to tackle problems with smaller signal-to-noise-ratio. On the other hand, it demands larger storage and more intensive computations, forcing the data science community to strive for efficient algorithms which can run in parallel in a distributed system. Many existing algorithms and methods (e.g., support vector machines) that are known to work well in small data scenarios do not scale well in a big data setting \citep{cortes1995support, forero2010consensus, wang2012distributed}. Recently, there has been an increasing amount of research interest in meta-algorithms, which can extend algorithms that are difficult to parallelize into distributed algorithms \citep{bottou2010large,zinkevich2010parallelized}, and ideas that resemble the divide-and-conquer algorithm \citep{kleiner2014scalable, jordan2018communication}.

At the same time, there is a class of algorithms which are trivially parallelizable and therefore can downsize big data problems into smaller ones. The key idea behind them, which dates back to \citet{fisher1922mathematical}, is to summarize the original data set by a low-dimensional vector of summary statistics, which can often be computed in a parallel and distributed way. For example, to estimate the mean and variance of a normal distribution from independent and identically distributed (i.i.d.) observations, we only need to obtain their sum and sum of squares, which are the corresponding summary statistics\footnote{In fact, they are \emph{sufficient statistics} \citep{casella2002statistical}, i.e., they can represent the original data-set perfectly without losing any information.} and can be trivially computed in a distributed fashion. In data-driven businesses such as information technology, these summary statistics are often referred to as \emph{metrics}, and used for measuring and monitoring key performance indicators \citep{dmitriev2016measuring, Deng:2016b}. In practice, it is often the changes or differences between metrics, rather than  measurements at the most granular level, that are of greater interest. In the context of randomized controlled experimentation (or A/B testing), inferring the changes of metrics can establish causality \citep{abScale,googlesurvey,xu2015infrastructure}, e.g., whether a new user interface design causes longer view times and more clicks.

\subsection{Central limit theorem and Delta method}

We advocate directly modeling the metrics rather than the original data-set. When analyzing data at the most granular level (e.g., user), we need some basic assumptions of the underlying probabilistic model, such as i.i.d. observations. When looking at the metrics level, we also need to know their joint distribution. This is where the blessing of big data comes into play. Given large sample sizes, the metrics often possess desirable asymptotic properties due to the \emph{central limit theorem} \citep{VanderVaart2000}. To ensure that the paper is self-contained, we first review the central limit theorem in its most well-known form. Let $X_1, \ldots, X_n$ be $n$ i.i.d. observations with finite mean $\mu$ and variance $\sigma^2 > 0.$ We let $\bar X$ denote the sample average, then as the sample size $n \rightarrow \infty,$ in distribution
$$
\sqrt{n}(\bar X -\mu) /\sigma \rightarrow N(0, 1).
$$
A common application of the central limit theorem is to construct the $100(1-\alpha)\%$ confidence interval of $\mu$ as
$
\bar{X} \pm z_{\alpha/2} \sigma/\sqrt{n},
$
where $z_{\alpha/2}$ is the $(\alpha / 2)$th quantile for $N(0, 1).$ This is arguably one of the most influential results of asymptotic statistics used in almost all scientific fields whenever an estimate with error bars is presented.

While an influential result, the central limit theorem in its basic form only applies to the average of i.i.d. random variables, and in practice our metrics are often more complex. To enjoy the blessing of big data, we employ the Delta method, which extends the normal approximations from the central limit theorem broadly. For illustration we only review the uni-variate case. For any random variable $T_n$ (the subscript indicates its dependency on $n,$ e.g., sample average) and constant $\theta$ such that
$
\sqrt{n}(T_n - \theta) \rightarrow N(0, 1)
$
in distribution as $n \rightarrow \infty,$ the Delta method allows us to extend its asymptotic normality to any continuous transformation $\phi(T_n).$ To be more specific, by using the fact that
$T_n-\theta = O(1/\sqrt{n})$ and the first order Taylor expansion \citep{rudin1964principles}
\begin{equation}
\label{eq:dm}
\phi(T_n)-\phi(\theta) = \phi'(\theta)(T_n-\theta) + O\{(T_n-\theta)^2\},
\end{equation}
we have in distribution
$$
\sqrt{n}
\left\{
\phi(T_n)-\phi(\theta)
\right\}
\rightarrow N 
\left\{ 
0, \phi'(\theta)^2 
\right\}.
$$
This is the Delta method. Without relying on any assumptions other than ``big data,'' the Delta method is \emph{general}. It is also \emph{memorable} -- anyone with basic knowledge of calculus can derive it. Moreover, the calculation is trivially \emph{parallelizable} and can be easily implemented in a distributed system. Nevertheless, although conceptually and theoretically straightforward, the \emph{practical} difficulty is to find the right ``link'' function $\phi$ that transforms the simple average to our desired metric. Because of different attributes of the metrics, such as the underlying data generating process and aggregation levels, the process of discovering the corresponding transformation can be challenging. However, unfortunately, although various applications of the Delta method have previously appeared in the data mining literature \citep{deng2013cuped,abScale,googlesurvey,kohavi2010}, the method itself and the discovery of $\phi$ were often deemed technical details and only briefly mentioned or relegated to appendices. Motivated by this gap, we aim to provide a practical guide that highlights the effectiveness and importance of the Delta method, hoping to help fellow data scientists, developers and applied researchers conduct trustworthy metric analytics.

\subsection{Scope and contributions}

As a practical guide, this paper presents four applications of the Delta method in real-life scenarios, all of which have been deployed in Microsoft's online A/B testing platform ExP  \citep{abScale,kohavi2009online} and employed to analyze experimental results on a daily basis: 
\begin{itemize}

\item In Section \ref{sec:pDelta}, we derive the asymptotic distributions of ratio metrics\footnote{Pedantically, ratio of two measurements of the same metric, from the treatment and control groups.}. Compared with standard approaches by \citet{fieller1940biological,fieller1954some}, the Delta method provides a much simpler and yet almost equally accurate and effective solution; 

\item In Section \ref{sec:cluster}, we analyze cluster randomized experiments, where the Delta method offers an efficient alternative algorithm to standard statistical machinery known as the mixed effect model \citep{bates2014lme4,dagelman}, and provides unbiased estimates; 

\item In Section \ref{sec:quantile}, by extending the Delta method to outer confidence intervals \citep{meyer1987outer}, we propose a \emph{novel} hybrid method to construct confidence intervals for quantile metrics with almost no extra computational cost\footnote{Our procedure computes two more quantiles for confidence interval. Since the main cost of quantile computing is sorting the data, computing three and one quantiles cost almost the same.}. Unlike most existing methods, our proposal does not require repeated simulations as in bootstrap, nor does it require estimating an unknown density function, which itself is often a theoretically challenging and computationally intensive task, and has been a center of criticism \citep{brown1983estimation};  

\item In Section \ref{sec:within}, we handle missing data in within-subject studies by combining the Delta method with data augmentation. We demonstrate the effectiveness of ``big data small problem'' approach when directly modeling metrics. Comparing to alternative methods that need to put up a model for individual subjects, our method requires less model assumptions.

\end{itemize}

The main purpose of this paper is to promote the Delta method as a \emph{general}, \emph{memorable} and \emph{efficient} tool for metric analytics. In particular, our contributions to the existing data mining literature include:
\begin{itemize}

\item Practical guidance for scenarios such as inferring ratio metrics, where the Delta method offers scalable and easier-to-implement solutions; 

\item A novel and computationally efficient solution to estimate the sampling variance of quantile metrics;

\item A novel data augmentation technique that employs the Delta method to model metrics resulting from within-subject or longitudinal studies with unspecified missing data patterns.
    
\end{itemize}
For reproducibility and knowledge transfer, we provide all relevant computer programs at https://aka.ms/exp/deltamethod.

\section{Inferring percent changes}\label{sec:pDelta}

\subsection{Percent change and Fieller interval}

Measuring \emph{change} is a common theme in applied data science. In online A/B testing \citep{abScale,googlesurvey,Deng:2016b,kohavi2007practical,DmitrievDirtyDozen}, we estimate the average treatment effect (ATE) by the difference of the same metric measured from treatment and control groups, respectively. In time series analyses and longitudinal studies, we often track a metric over time and monitor changes between different time points. For illustration, let $X_1, \ldots, X_n$ be i.i.d. observations from the control group with mean $\mu_x$ and variance $\sigma_x^2,$  $Y_1, \ldots, Y_n$ i.i.d. observations from the treatment group with mean $\mu_y$ and variance $\sigma_y^2,$ and $\sigma_{xy}$ the covariance\footnote{For A/B testing where the treatment and control groups are independently sampled from a super population, $\sigma_{xy} = 0.$} between $X_i$'s and $Y_j$'s. Let $\bar X = \sum_{i=1}^n X_i / n$ and $\bar Y = \sum_{i=1}^n Y_i / n$ be two measurements of the same metric from the treatment and control groups, respectively, and their difference $\hat \Delta = \bar Y - \bar X$ is an unbiased estimate of the ATE $\Delta = \mu_y - \mu_x.$ Because both $\bar X$ and $\bar Y$ are approximately normally distributed, their difference $\hat \Delta$ also follows an approximate normal distribution with mean $\Delta$ and variance
$$
\var( \hat \Delta) = ( \sigma_y^2 + \sigma_x^2 -2\sigma_{xy} ) / n.
$$
Consequently, the well-known 100(1-$\alpha$)\% confidence interval of $\Delta$ is 
$
\hat \Delta \pm z_\alpha\times \widehat{\var}(\hat \Delta),
$
where $\widehat{\var}(\hat \Delta)$ is the finite-sample analogue of $\var(\hat \Delta),$ and can be computed using the sample variances and covariance of the treatment and control observations, denoted as $s_y^2,$ $s_x^2$ and $s_{xy}$ respectively.

In practice, however, \emph{absolute} differences as defined above are often hard to interpret because they are not scale-invariant. Instead, we focus on the \emph{relative} difference or percent change $\Delta\% = ( \mu_y-\mu_x ) / \mu_x,$ estimated by 
$\hat \Delta\% = (\bar Y - \bar X) / \bar X.$ The key problem of this section is constructing a 100(1-$\alpha$)\% confidence interval for $\hat \Delta\%$. For this classic problem, \citet{fieller1940biological, fieller1954some} seems to be the first to provide a solution. To be specific, let $t_r(\alpha)$ denote the $(1-\alpha)$th quantile for the $t-$distribution with degree of freedom $r,$ and 
$
g = n s_x^2 t_{\alpha / 2}^2(r) / \bar X^2,
$
then Fieller's interval of $\Delta\%$ is
\begin{equation}
\label{eq:fieller}
\begin{split}
\frac{1}{1-g}
&\Bigg\{ 
\frac{\bar Y}{\bar X} - 1 
-
\frac{g s_{xy}}{s_x^2} \\ 
&\pm  
\frac{t_{\alpha/2}(r)}{\sqrt{n} \bar X}
\sqrt{
s_y^2 - 2\frac{\bar Y}{\bar X}s_{xy} + \frac{\bar Y^2}{\bar X^2}s_x^2 
- 
g 
\left( 
s_y^2 - \frac{s_{xy}^2}{s_x^2}
\right)
}
\Bigg\}
\end{split}
\end{equation}
Although widely considered as the standard approach for estimating variances of percent changes \citep{kohavi2009controlled}, deriving \eqref{eq:fieller} is cumbersome (see \citep{von2009geometric} for a modern revisit). The greater problem is that, even when given this formula, applying it often requires quite some effort. According to \eqref{eq:fieller} we need to not only estimate the sample variances and covariance, but also the parameter $g.$

\subsection{Delta method and Edgeworth correction}
\label{subsec:ratiometrics}

The Delta method provides a more intuitive alternative solution. Although they can be found in classic textbooks such as \citep{casella2002statistical}, this paper (as a practical guide) still provides all the relevant technical details. We let $T_n = (\bar Y, \bar X)$, $\theta = (\mu_y, \mu_x)$ and $\phi(x,y) = y / x.$ A multi-variate analogue of \eqref{eq:dm} suggests that
$
\phi(T_n)-\phi(\theta) \approx \nabla\phi(\theta) \cdot (T_n-\theta),
$
which implies that
\begin{equation}
\label{eq:taylor-expansion}
\frac{Y}{X} - \frac{\mu_y}{\mu_x} 
\approx 
\frac{1}{\mu_x}(\bar Y - \mu_y) - \frac{\mu_y}{\mu_x^2}(\bar X - \mu_x)
\end{equation}
For $i = 1, \ldots, n,$ let
$
W_i =  Y_i / \mu_x - \mu_y X_i / \mu_x^2,
$
which are also i.i.d. observations. Consequently, we can re-write \eqref{eq:taylor-expansion} as
$
\bar Y / \bar X - \mu_y / \mu_x 
\approx 
\sum_{i=1}^n  W_i / n,
$
leading to the Delta method based confidence interval
\begin{equation}
\label{eq:Deltaratio}
\underbrace{
\vphantom
{
\frac{z_{\alpha/2}}{\sqrt{n}\bar X} 
\sqrt{
s_y^2 
- 2 \frac{\bar Y}{\bar X}s_{xy} 
+ \frac{\bar Y^2}{\bar X^2}s_x^2
}
}
\frac{\bar Y}{\bar X} - 1 
}_{\textrm{point estimate}}
\pm 
\underbrace{
\frac{z_{\alpha/2}}{\sqrt{n}\bar X} 
\sqrt{
s_y^2 
- 2 \frac{\bar Y}{\bar X}s_{xy} 
+ \frac{\bar Y^2}{\bar X^2}s_x^2
}
}_{\textrm{uncertainty quantification}}
\end{equation}
\eqref{eq:Deltaratio} is easier to implement than \eqref{eq:fieller}, and is in fact the limit of \eqref{eq:fieller} in the large sample case, because as $n \rightarrow \infty,$ we have
$
t_{\alpha/2}(r) \to z_{\alpha/2},
$
$
g \to 0,
$
and
$
\var(\bar X)\sim O(1/n).
$ 
Although Fieller's confidence interval can be more accurate for small samples \citep{hirschberg2010geometric}, this benefit appears rather limited for big data. Moreover, the Delta method can also be easily extended for a better approximation by using Edgeworth expansion \citep{hall2013bootstrap,boos2000large}. To be more specific, \eqref{eq:taylor-expansion} suggests that in distribution
$
\sqrt{n}
\left(
\bar Y / \bar X - \mu_y / \mu_x 
\right)
/ \sigma_w
\rightarrow
\nu,
$
whose cumulative distribution function
$$
F(t) = \Phi(t) - 6n^{-1/2}\kappa_w(t^2-1)\phi(t) 
$$
contains a correction term in addition to $\Phi(t),$ the cumulative distribution function of the standard normal, and $\kappa_w$, the skewness of the $W_i$'s. By simply replacing the terms ``$\pm z_{\alpha/2}$'' in \eqref{eq:Deltaratio} with $\nu_{\alpha / 2}$ and $\nu_{1 - \alpha / 2},$ the $(\alpha / 2)$th and $(1 - \alpha / 2)$th quantiles of $\nu,$ respectively, we obtain the corresponding Edgeworth expansion based confidence interval. Finally, we can add a second-order bias correction term 
$
( \bar Y s_x^2 / \bar X - s_{xy}) / (n \bar X)^2
$
to the point estimate in \eqref{eq:Deltaratio}; the same correct term can be applied to the Edgeworth expansion based interval.

\subsection{Numerical examples}\label{sec:empirical1}

To illustrate the performance of Fieller's interval in \eqref{eq:fieller}, the Delta method interval in \eqref{eq:Deltaratio}, and the Edgeworth interval with and without bias correction under different scenarios, we let the sample size $n = 20, 50, 200, 2000.$ For each fixed $n,$ we assume that the treatment and control groups are independent, and consider three simulation models for i.i.d. experimental units $i=1, \ldots, n:$
\begin{enumerate}

\item $\text{Normal:} \; X_i \sim N(\mu = 1, \sigma = 0.1),
\;
Y_i \sim N(\mu = 1.1, \sigma = 0.1);$

\item $\text{Poisson:} \; X_i \sim \mathrm{Pois}(\lambda = 1),
\;
Y_i \sim \mathrm{Pois}(\lambda = 1.1);$

\item $\text{Bernoulli:} \; X_i \sim \mathrm{Bern}(p = 0.5),
\;
Y_i \sim \mathrm{Bern}(p = 0.6).$

\end{enumerate}
The above models aim to mimic the behaviors of our most common metrics, discrete or continuous. For each case, we repeatedly sample $M = 10,000$ data sets, and for each data set we construct Fieller's and the Delta method intervals, respectively, and then add correction to the Delta method result. We also construct the Edgeworth expansion interval without and with the bias correction. 
\begin{table}[ht]
\centering\caption{\label{tab:coverage} \footnotesize Simulated examples: The first two columns contain simulation models and sample sizes. The next five columns present the coverage rates of various 95\% confidence intervals -- Fieller's, Delta method based (w/o and w/ bias correction) and Edgeworth expansion based (w/o and w/ bias correction).}

\resizebox{\linewidth}{!}{\begin{tabular}{lrrrrrr}
\toprule
Method & n & Fieller & Delta & Delta(BC) & Edgeworth & Edgeworth(BC)\\
\midrule
Normal & 20 & 0.9563 & 0.9421 & 0.9422 & 0.9426 & 0.9426\\
Normal & 50 & 0.9529 & 0.9477 & 0.9477 & 0.9478 & 0.9477\\
Normal & 200 & 0.9505 & 0.9490 & 0.9491 & 0.9490 & 0.9490\\
Normal & 2000 & 0.9504 & 0.9503 & 0.9503 & 0.9503 & 0.9503\\
\addlinespace
Poisson & 20 & 0.9400 & 0.9322 & 0.9370 & 0.9341 & 0.9396\\
Poisson & 50 & 0.9481 & 0.9448 & 0.9464 & 0.9464 & 0.9478\\
Poisson & 200 & 0.9500 & 0.9491 & 0.9493 & 0.9496 & 0.9498\\
Poisson & 2000 & 0.9494 & 0.9494 & 0.9495 & 0.9494 & 0.9495\\
\addlinespace
Bernoulli & 20 & 0.9539 & 0.9403 & 0.9490 & 0.9476 & 0.9521\\
Bernoulli & 50 & 0.9547 & 0.9507 & 0.9484 & 0.9513 & 0.9539\\
Bernoulli & 200 & 0.9525 & 0.9513 & 0.9509 & 0.9517 & 0.9513\\
Bernoulli & 2000 & 0.9502 & 0.9500 & 0.9499 & 0.9501 & 0.9500\\
\bottomrule
\end{tabular}}
\end{table}

We report the corresponding coverage rates in Table \ref{tab:coverage}, from which we can draw several conclusions. For big data ($n \ge 200$), all methods achieve nominal (i.e., $\approx 95\%$) coverage rates for all simulation models. For small data ($n \le 50$), although Fieller's interval seems more accurate for some simulation models (e.g., normal), other methods perform comparably, especially after the bias correction. For simplicity in implementation and transparency in applications, we recommend Algorithm \ref{algo:delta-bc}, which uses the Delta method based interval \eqref{eq:Deltaratio} with the bias correction. 
\begin{algorithm}
\small
   \caption{\small Confidence interval for ratio: Delta method + bias correction}
    \begin{algorithmic}[1]
      \Function{deltaci}{$X = X_1, \ldots, X_n; Y_1, \ldots, Y_n;$ $\alpha = 0.05$}

        \State $\bar X = \textrm{mean}(X);$ $\bar Y = \textrm{mean}(Y);$
        \State $s_x^2 = \textrm{var}(X);$ $s_y^2 = \textrm{var}(Y);$ $s_{xy} = \textrm{cov}(X,Y);$ 
        
        \State bc =  $\bar Y / \bar X^3 \times s_x^2 / n - 1/\bar X^2 \times s_{xy}/ n$ \Comment{bias correction term}
        \State pest = $\bar Y / \bar X - 1$ + bc \Comment{point estimate}
        \State vest = $s_y^2 / \bar X^2 - 2 \times \bar Y / \bar X^3 * s_{xt} + \bar Y^2 / \bar X^4 * s_x^2$
        
        \State return: $\textrm{pest} \pm z_{1-\alpha/2} \times \sqrt{\textrm{vest} / n}$ \Comment{$100(1-\alpha)$ confidence interval}
       \EndFunction

\end{algorithmic}
\label{algo:delta-bc}
\end{algorithm}

\section{Decoding cluster randomization}
\label{sec:cluster}

\subsection{The variance estimation problem}

Two key concepts in a typical A/B test are the \emph{randomization unit} -- the granularity level where sampling or randomization is performed, and the \emph{analysis unit} -- the aggregation level of metric computation. Analysis is straightforward when the randomization and analysis units agree \citep{Deng:2017}, e.g., when randomizing by user while also computing the average revenue per user. However, often the randomization unit is a \emph{cluster} of analysis units (it cannot be more granular than the analysis unit,  otherwise the analysis unit would contain observations under both treatment and control, nullifying the purpose of differentiating the two groups). Such cases, sometimes referred to as cluster randomized experiments in the econometrics and statistics literature \citep{athey2017econometrics,klar2001current}, are quite common in practice, e.g., enterprise policy prohibiting users within the same organization from different experiences, or the need to reduce bias in the presence of network interference \citep{backstrom2011network,eckles2017design,gui2015network}. Perhaps more ubiquitously, for the same experiment we usually have metrics with different analysis units. For example, to meet different business needs, most user-randomized experiments run by ExP contain both user-level and page-level metrics.

We consider two average metrics\footnote{Pedantically, they are two measurements of the same metric. We often use metric to refer to both the metric itself (e.g. revenue-per-user) and measurements of the metric (e.g. revenue-per-user of the treatment group) and this distinction would be clear in the context.} of the treatment and control groups, assumed to be independent. Without loss of generality, we only focus on the treatment group with $K$ clusters. For $i = 1, \ldots, K,$ the $i$th cluster contains $N_i$ analysis unit level observations $Y_{ij}$ ($j=1,\dots,N_i$). Then the corresponding average metric is
$
\bar{Y} = \sum_{i,j}Y_{ij} \big/ \sum_i N_i.
$
We assume that within each cluster the observations $Y_{ij}$'s are i.i.d. with mean $\mu_i$ and variance $\sigma_i^2$, and across clusters $(\mu_i,\sigma_i,N_i)$ are also i.i.d.

\subsection{Existing and Delta method based solutions}

$\bar{Y}$ is not an average of i.i.d. random variables, and the crux of our analysis is to estimate its variance. Under strict assumptions, closed-form solutions for this problem exist \cite{donner1987statistical,klar2001current}. For example, when $N_i = m$ and $\sigma_i^2 = \sigma^2$ for all $i,$
\begin{equation}
\label{eq:donner}
    \var(\bar{Y}) = \frac{\sigma^2+\tau^2}{Km}\{1+(m-1)\rho\},
\end{equation}
where $\tau^2 = \var(\mu_i)$ is the \emph{between-cluster} variance and $\rho = \tau^2 / (\sigma^2+\tau^2)$ is the \emph{coefficient of intra-cluster correlation}, which quantifies the contribution of between-cluster variance to the total variance. To facilitate understanding of the variance formula \eqref{eq:donner}, two extreme cases are worth mentioning:
\begin{enumerate}

\item If $\sigma=0,$ then for each $i=1,\dots,K$ and all $j = 1, \ldots, N_i,$ $Y_{ij} = \mu_i.$ In this case, $\rho=1$ and $\var(\bar{Y})=\tau^2/K;$

\item If $\tau=0$, then $\mu_i = \mu$ for all $i=1, \ldots, K,$ and therefore the observations $Y_{ij}$'s are in fact i.i.d. In this case, $\rho=0$ and \eqref{eq:donner} reduces to $\var(\bar{Y})=\sigma^2/(Km)$.

\end{enumerate}
However, unfortunately, although theoretically sound, \eqref{eq:donner} has only limited practical value because the assumptions it makes are unrealistic. In reality, the cluster sizes $N_i$ and distributions of $Y_{ij}$ for each cluster $i$ are different, which means that $\mu_i$ and $\sigma_i^2$ are different.

Another common approach is the mixed effect model, also known as multi-level/hierarchical regression \citep{dagelman}, where $Y_{ij}$ depends on $\mu_i$ and $\sigma_i^2,$ while the parameters themselves follow a ``higher order'' distribution. Under this setting, we can infer the treatment effect as the ``fixed'' effect for the treatment indicator term\footnote{Y $\sim$ Treatment + (1|User) in \emph{lme4} notation. A detailed discussion is beyond the scope of this paper; see \citet{dagelman,bates2014fitting}.}. \emph{Stan} \citep{carpenter2016stan} offers a Markov Chain Monte Carlo (MCMC) implementation aiming to infer the posterior distribution of those parameters, but this needs significant computational effort for big data. Moreover, the estimated ATE, i.e., the coefficient for the treatment assignment indicator, is for the randomization unit (i.e., cluster) but not the analysis unit level, because it treats all clusters with equal weights and can be viewed as the effect on the double average $\sum_i (\sum_j Y_{ij} / N_j) / K,$ which is usually different than the population average $\bar Y$ \citep{Deng:2016b}. This distinction doesn't make a systematic difference when effects across clusters are homogeneous. However, in practice the treatment effects are often heterogeneous, and using mixed effect model estimates without further adjustment steps could lead to severe biases.

On the contrary, the Delta method solves the problem directly from the metric definition. Re-write $\bar Y$ into 
$
\sum_i (\sum_j Y_{ij}) / \sum_i N_i.
$
Let $S_i = \sum_j Y_{ij}$, and divide both the numerator and denominator by $K$, 
$$
\bar{Y} = \frac{\sum_i S_i / K}{\sum_i N_i /K} = {\bar S}/{\bar N}.
$$
Albeit not an average of i.i.d. random variables, $\bar{Y}$ is a ratio of two averages of i.i.d. randomization unit level quantities \citep{Deng:2017}. Therefore, by following \eqref{eq:taylor-expansion} in Section \ref{subsec:ratiometrics},
\begin{align}
\label{eq:varratio2}
\var(\bar{Y}) 
& \approx \frac{1}{K\mu_N^2}
\left(
\sigma_S^2 - 2\frac{\mu_S}{\mu_N} \sigma_{SN} + \frac{\mu_S^2}{\mu_N^2} \sigma_N^2
\right).
\end{align}
Therefore, the variance of $\bar Y$ depends on the variance of a centered version of $S_i$ (by a multiple of $N_i$, not the constant $\e(S_i)$ as typically done). Intuitively, both the variance of the cluster size $N_i,$ and of the within-cluster sum of observations $S_i = \sum_j Y_{ij},$ contribute to \eqref{eq:varratio2}. In particular, $\sigma_N^2 = \var{N_i}$ is an important contributor of the variance of $\bar Y,$ leading to a practical recommendation for the experimenters -- it is desirable to make the cluster sizes homogeneous; otherwise it can be difficult to detect small treatment effects due to low statistical power.

\subsection{Numerical examples}\label{sec:empirical2}

Because the coverage property of \eqref{eq:varratio2} has been extensively covered in Section ~\ref{sec:empirical1}, we only focus on comparing it with the mixed effect model here. We first describe the data generating process, which consists of a two-level hierarchy as described in the previous section. First, at the randomization unit level, let the total number of clusters $K = 1000.$ To mimic cluster heterogeneity, we divide clusters into three categories: small, medium and large. We generate the numbers of clusters in the three categories by the following multinomial distribution:
$$
(M_1, M_2, M_3)^\prime 
\sim \textrm{Multi-nomial}
\{
n = K; p = (1/3, 1/2, 1/6)
\}.
$$
For the $i$th cluster, depending on which category it belongs to, we generate $N_i, \mu_i$ and $\sigma_i$ in the following way\footnote{The positive correlation between $\mu_i$ and $N_i$ is not important, and reader can try out code with different configuration.}:
\begin{itemize}
    
    \item Small: $N_i \sim \textrm{Poisson}(2),$ $\mu_i \sim N(\mu = 0.3, \sigma_i = 0.05);$ 
    
    \item Medium: $N_i \sim \textrm{Poisson}(5),$ $\mu_i \sim N(\mu = 0.5, \sigma_i = 0.1);$
    
    \item High: $N_i \sim \textrm{Poisson}(30),$ $\mu_i \sim N(\mu = 0.8, \sigma_i = 0.05);$

\end{itemize}
Second, for each fixed $i,$ let
$
Y_{ij} \sim \textrm{Bernoulli}(p = \mu_i)
$
for all
$
j = 1, \ldots, N_i.
$
This choice is motivated by binary metrics such as page-click-rate, and because of it we can determine the ground truth $\e(\bar Y) = 0.667$ by computing the weighted average of $\mu_i$ weighted by the cluster sizes and the mixture of small, medium and large clusters.

Our goal is to infer $\e(\bar Y)$ and we compare the following three methods: 
\begin{enumerate}

\item A naive estimator $\bar{Y},$ pretending all observations $Y_{ij}$ are i.i.d;

\item Fitting a mixed effect model with a cluster random effect $\mu_i;$ 

\item Using the metric $\bar{Y}$ as in the first method, but using the Delta method for variance estimation. 

\end{enumerate}
Based on the aforementioned data generating mechanism, we repeatedly and independently generate 1000 data sets. For each data set, we compute the point and variance estimates of $\e(\bar Y)$ using the naive, mixed effect, and delta methods. We then compute empirical variances for the three estimators, and compare them to the average of estimated variances. We report the results in Table ~\ref{tab:clustered}.
\begin{table}[h]
\caption{\footnotesize Simulated examples: The first three columns contain the chosen method, the true value of $\e(\bar Y)$ and the true standard deviation of the corresponding methods. The last two columns contain the point estimates and average estimated standard errors.}
\centering
\resizebox{\linewidth}{!}{\begin{tabular}[t]{lrrrr}
\toprule
Method  & Ground Truth & SD(True) & Estimate & Avg. SE(Model)\\
\midrule
Naive & 0.667 & 0.00895 & 0.667 & 0.00522\\
Mixed effect & 0.667 & 0.00977 & 0.547 & 0.00956\\
Delta method & 0.667 & 0.00895 & 0.667 & 0.00908\\
\bottomrule
\end{tabular}}
\label{tab:clustered}
\end{table}
\enskip

Not surprisingly, the naive method under-estimates the true variance -- we had treated the correlated observations as independent. Both the Delta method and the mixed effect model produced satisfactory variance estimates. However, although both the naive and the Delta method correctly estimated $\e(Y)$, the mixed effect estimator is severely biased. This shouldn't be a big surprise if we look deeper into the model
$
Y_{ij} = \alpha + \beta_i +\epsilon_{ij}
$
and
$
\e(\epsilon_{ij})=0,
$
where the random effects $\beta_i$ are centered so $\e(\beta_i)=0$. The sum of the intercept terms $\alpha$ and $\beta_i$ stands for the per-cluster mean $\mu_i$, and $\alpha$ represents the average of per-cluster mean, where we compute the mean within each cluster first, and then average over clusters. This is different from the metrics defined as simple average of $Y_{ij}$ in the way that in the former all clusters are equally weighted and in the latter case bigger clusters have more weight. The two definitions will be the same if and only if either there is no heterogeneity, i.e. per-cluster means $\mu_i$ are all the same, or all clusters have the same size. We can still use the mixed effect model to get a unbiased estimate. This requires us to first estimate every $\beta_i$ (thus $\mu_i$), and then compute $(\alpha+\beta_i)N_i / \sum_i N_i$ by applying the correct weight $N_i$.  The mixed effect model with the above formula gave a new estimate $0.662$, much closer to the ground truth. Unfortunately, it is still hard to get the variance of this new estimator. 

In this study we didn't consider the treatment effect. In ATE estimation, the mixed effect model will similarly result in a biased estimate for the ATE for the same reason, as long as per-cluster treatment effects vary and cluster sizes are different. The fact that the mixed effect model provides a double average type estimate and the Delta method estimates the ``population'' mean is analogous to the comparison of the mixed effect model with GEE (generalized estimating equations) \citep{liang1986longitudinal}. In fact, in the Gaussian case, the Delta method can be seen as the ultimate simplification of GEE's sandwich variance estimator after summarizing data points into sufficient statistics. But the derivation of GEE is much more involved than the central limit theorem, while we can explain the Delta method in a few lines and it is not only more \emph{memorable} but also provides more insights in \eqref{eq:varratio2}.

\section{Efficient variance estimation for quantile metrics}\label{sec:quantile}

\subsection{Sample quantiles and their asymptotics}

Although the vast majority of metrics are averages of user telemetry data, quantile metrics form another category that is widely used to focus on the tails of distributions. In particular, this is often the case for performance measurements, where we not only care about an average user's experience, but even more so about those that suffer from the slowest responses. Within the web performance community, quantiles (of, for example, page loading time) at 75\%, 95\% or 99\% often take the spotlight. In addition, the 50\% quantile (median) is sometimes used to replace the mean, because it is more robust to outlier observations (e.g., errors in instrumentation). This section focuses on estimating the variances of quantile estimates. 

Suppose we have $n$ i.i.d. observations $X_1,\dots,X_n,$ generated by a cumulative distribution function $F(x) = P(X\le x)$ and a density function $f(x)$\footnote{We do not consider cases when $X$ has discrete mass, and $F$ will have jumps. In this case the quantile can take many values and is not well defined. In practice this case can be seen as continuous case with some discrete correction.}. The theoretical $p$th quantile for the distribution $F$ is defined as $F^{-1}(p)$. Let 
$
X_{(1)}, \dots, X_{(n)} 
$ 
be the ascending \emph{ordering} of the original observations. The sample quantile at $p$ is $X_{(np)}$ if $np$ is an integer. Otherwise, let $\lfloor np\rfloor$ be the floor of $np$, then the sample quantile can be defined as any number between $X_{(\lfloor np\rfloor)}$ and $X_{(\lfloor np\rfloor +1)}$ or a linear interpolation of the two\footnote{When p is 0.5, the 50\% quantile, or median, is often defined as the average of the middle two numbers if we have even number of observations.}. For simplicity here we use $X_{(\lfloor np\rfloor)},$ which will not affect any asymptotic results. It is a well-known fact that, if $X_1, \ldots, X_n$ are i.i.d. observations, following the central limit theorem and a rather straightforward application of the Delta method, the sample quantile is approximately normal \citep{casella2002statistical, VanderVaart2000}: 
\begin{equation}
\label{eq:quantileclt}
\sqrt{n}
\left\{
X_{\lfloor np\rfloor} - F^{-1}(p)
\right\}
\rightarrow 
N
\left[
0, \frac{\sigma^2}{f \left\{F^{-1}(p)\right\}^2}
\right],
\end{equation}
where 
$
\sigma^2 = p(1-p).
$
However, unfortunately, in practice we rarely have i.i.d. observations. A common scenario is in search engine performance telemetry, where we receive an observation (e.g., page loading time) for each server request or page-view, while randomization is done at a higher level such as device or user. This is the same situation we have seen in Section~\ref{sec:cluster}, where $X_i$ are clustered. To simplify future notations, we let 
$
Y_i = I \{ X_i \le F^{-1}(p) \}, 
$
where $I$ is the indicator function. Then \eqref{eq:varratio2} can be used to compute $\var(\bar Y)$, and \eqref{eq:quantileclt} holds in the clustered case with $\sigma^2 = n \var(\bar Y)$. This generalizes the i.i.d. case where $n\var(\bar Y) = p(1-p)$. Note that the Delta method is  instrumental in proving \eqref{eq:quantileclt} itself, but a rigorous proof involves a rather technical last step that is beyond our scope. A formal proof can be found in \citep{VanderVaart2000}.

\subsection{A Delta method solution to a practical issue}

Although theoretically sound, the difficulty of applying \eqref{eq:quantileclt} in practice lies in the denominator 
$
f \{ F^{-1}(p) \},
$
whose computation requires the \emph{unknown} density function $f$ at the \emph{unknown} quantile $F^{-1}(p).$ A common approach is to estimate $f$ from the observed $X_i$ using non-parametric methods such as kernel density estimation \citep{allofstat}. However, any non-parametric density estimation method is trading off between bias and variance. To reduce variance, more aggressive smoothing and hence larger bias need to be introduced to the procedure. This issue is less critical for quantiles at the body of the distribution, e.g. median, where density is high and more data exists around the quantile to make the variance smaller. As we move to the tail, e.g. 90\%, 95\% or 99\%, however, the noise of the density estimation gets bigger, so we have to introduce more smoothing and more bias. Because the density shows up in the denominator and density in the tail often decays to $0$, a small bias in estimated density can lead to a big bias for the estimated variance (\citet{brown1983estimation} raised similar criticisms with their simulation study). A second approach is to bootstrap, re-sampling the whole dataset many times and computing quantiles repeatedly. Unlike an average, computing quantiles requires sorting, and sorting in distributed systems (data is distributed in the network) requires data shuffling between nodes, which incurs costly network I/O. Thus, bootstrap works well for small scale data but tends to be too expensive in large scale in its original form (efficient bootstrap on massive data is a research area of its own \citep{kleiner2014scalable}).

An alternative method without the necessity for density estimation is more desirable, especially from a more practical perspective. One such method is called outer confidence interval (outer CI) \citep{krewski1976distribution,meyer1987outer}, which produces a closed-form formula for quantile confidence intervals using combinatorial arguments. Recall that
$
Y_i = I \{X_i \le F^{-1}(p) \}
$ 
and $\sum Y_i$ is the count of observations no greater than the quantile. In the aforementioned i.i.d. case, $\sum Y_i$ follows a binomial distribution. Consequently, when $n$ is large
$
\sqrt{n}(\bar Y-p) \approx N(0, \sigma^2)
$
where $\sigma^2 = p(1-p)$. If the quantile value 
$
F^{-1}(p) \in [X_{(r)},X_{(r+1)}),
$ 
then $\bar Y = r/n$. The above equation can be inverted into a $100(1-\alpha)\%$ confidence interval for $r/n:$ $p \pm z_{\alpha/2}\sigma/\sqrt{n}$. This means with 95\% probability the true percentile is between the lower rank $L = n(p-z_{\alpha/2}\sigma/\sqrt{n})$ and upper rank $U = n(p+z_{\alpha/2}\sigma/\sqrt{n})+1$!

The traditional outer CI depends on $X_i$ being i.i.d. But when there are clusters, $\sigma^2/n$ instead of being $p(1-p)/n$ simply takes a different formula \eqref{eq:varratio2} by the Delta method and the result above still holds. Hence the confidence interval for a quantile can be computed in the following steps:
\begin{enumerate}
    \item fetch the quantile $X_{(\lfloor np\rfloor)}$
    \item compute $Y_i = I \{X_i \le X_{(\lfloor np\rfloor)}$\}
    \item compute $\mu_S$, ${\mu_N}$, $\sigma_S^2$, $\sigma_{SN}$, $\sigma_N^2$
    \item compute $\sigma$ by setting $\sigma^2/n$ equal to the result of equation~\eqref{eq:varratio2}
    \item compute $L,U = n(p \pm z_{\alpha/2}\sigma)$ 
    \item fetch the two ranks $X_{(L)}$ and $X_{(U)}$
\end{enumerate}
We call this outer CI with pre-adjustment. This method reduces the complexity of computing a quantile and its confidence interval into a Delta method step and subsequently fetching three ``ntiles''. However, in this algorithm the ranks depends on $\sigma$, whose computation depends on the quantile estimates (more specifically the $Y_i$ requires a pass through the data after quantile is estimated). This means that this algorithm requires a first `ntile' retrieval, and then a pass through the data for $\sigma$ computation, and then another two `ntile' retrievals. Turns out, computing all three `ntiles' in one stage is much more efficient than splitting into two stages. This is because retrieving `ntiles' can be optimized in the following way: if we only need to fetch tail ranks, it is pointless to sort data that are not at the tail; we can use sketching algorithm to narrow down the possible range where our ranks reside and only sort in that range, making it even more efficient to retrieve multiple `ntiles' at once. Along this line of thoughts, to make the algorithm more efficient, we noticed that \eqref{eq:quantileclt} also implies that the change from $X_i$ being i.i.d. to clustered only requires an adjustment to the numerator $\sigma$, which is a simple re-scaling step, and the correction factor does not depend on the unknown density function $f$ in the denominator. If the outer CI were to provide a good confidence interval in i.i.d. case, a re-scaled outer CI with the same correction term should also work for the clustered case, at least when $n$ is large. This leads to the outer CI with post-adjustment algorithm:
\begin{enumerate}
    \item compute $L,U = n(p \pm z_{\alpha/2}\sqrt{p(1-p)/n})$ 
    \item fetch $X_{(\lfloor np\rfloor)}$, $X_{(L)}$ and $X_{(U)}$
    \item compute $Y_i = I \{X_i \le X_{(\lfloor np\rfloor)}$\}
    \item compute $\mu_S$, ${\mu_N}$, $\sigma_S^2$, $\sigma_{SN}$, $\sigma_N^2$
    \item compute $\sigma$ by setting $\sigma^2/n$ equal to the result of equation~\eqref{eq:varratio2}
    \item compute the correction factor $\sigma / \sqrt{p(1-p)}$ and apply it to $X_{(L)}$ and $X_{(U)}$
\end{enumerate}
We implemented this latter method in ExP using Apache Spark \citep{zaharia2016apache} and SCOPE \citep{chaiken2008scope}.

\subsection{Numerical examples}
\label{sec:empirical3}

To test the validity and performance of the adjusted outer CI method, we compare its coverage to a standard non-parametric bootstrap ($N_B=1000$ replicates). The simulation setup consists of $N_u=100,\dots,10000$ users (clusters) with $N_i^u=1,\dots,10$ observations each ($N_i^u$ are uniformly distributed). Each observation is the sum of two i.i.d. random variables $X_i^u = X_i + X_u$, where $X_u$ is constant for each user. We consider two cases, one symmetric and one heavy-tailed distribution: 
\begin{itemize}

\item $\text{Normal:} \enskip X_i, X_u \stackrel{\mathrm{iid}}{\sim} \mathcal{N}(\mu = 0, \sigma = 1);$

\item $\text{Log-Normal:} \enskip X_i, X_u \stackrel{\mathrm{iid}}{\sim} \text{Log-normal}(\mu = 0, \sigma = 1).$

\end{itemize}
First, we find the "true" 95th percentile value of these distribution by computing its value for a very large sample ($N_u=10^7$). Second, we compute the confidence intervals for $M=10000$ simulation runs using bootstrap and outer CI with pre- and post-adjustment and compare their coverage estimates ($\approx$0.002 standard error), shown in Table ~\ref{tab:emp3coverage}. We found that when the sample contains 1000 or more clusters, all methods provide good coverage. Pre- and post-adjustment outer CI results are both very close to the much more computationally expensive bootstrap (in our un-optimized simulations, the outer CI method was $\approx$20 times faster than bootstrap). When the sample size was smaller than 1000 clusters, bootstrap was noticeably inferior to outer CI. For all sample sizes, pre-adjustment provided slightly larger coverage than post-adjustment, and this difference increased for smaller samples. In addition, because adjustment tends to result in increased confidence intervals, unadjusted ranks are more likely to have the same value as the quantile value, and thus post-adjustment is more likely to underestimate the variance in that case. In conclusion, post-adjustment outer CI works very well for large sample sizes $N_u \ge 1000$ and reasonably well for smaller samples, but has slightly inferior coverage compared to pre-adjustment. For big data, outer CI with post-adjustment is recommended due its efficiency, while for median to small sample sizes, outer CI with pre-adjustment is preferred if accuracy is paramount. 
\begin{table}[ht]
\centering
\caption{\footnotesize Simulated examples: The first two columns contain the simulated models and sample sizes. The last three columns contain the coverage rates for the Bootstrap, pre-adjusted and post-adjusted outer confidence intervals.}
\begin{tabular}{lrrrr}
  \hline
  Distribution & $N_u$ & Bootstrap & Outer CI  & Outer CI \\ 
  & & & (pre-adj.) & (post-adj.)\\
  \hline
  Normal & 100 & 0.9039 & 0.9465 & 0.9369\\ 
  Normal & 1000 & 0.9500 & 0.9549 & 0.9506\\ 
  Normal & 10000 & 0.9500 & 0.9500 & 0.9482\\
  \addlinespace
  Log-normal & 100 & 0.8551 & 0.9198 & 0.9049\\ 
  Log-normal & 1000 & 0.9403 & 0.9474 & 0.9421\\ 
  Log-normal & 10000 & 0.9458 & 0.9482 & 0.9479\\ 
   \hline
\end{tabular}
\label{tab:emp3coverage}
\end{table}

\section{Missing Data and Within-Subject Analyses}
\label{sec:within}

\subsection{Background}

Consider the case that we are tracking a metric over time. For example, businesses need to track key performance indicators like user engagement and revenue daily or weekly for a given audience. To simplify, let us say we are tracking weekly visits-per-user. Visits-per-user is defined as a simple average $\bar{X}_t, t = 1,\dots$ for each week. If there is a drop between $\bar{X}_t$ and $\bar{X}_{t-1}$, how do we set up an alert so that we can avoid triggering it due to random noise and control the false alarm rate? Looking at the variance, we see
$$\var(\bar{X}_{t}-\bar{X}_{t-1}) = \var(\bar{X}_t)+\var(\bar{X}_{t-1})-2\cov(\bar{X}_t,\bar{X}_{t-1})$$
and we need to have a good estimate of the covariance term because we know it is non-zero. 

Normally, estimating the covariance of two sample averages is trivial and the procedure is very similar to the estimation of the sample variance. But there is something special about this case --- missing data. Not everyone uses the product every week. For any online and offline metric, we are only able to define the metric using observed data. If for every user we observe its value in week $t-1$ and $t$, the covariance can be estimated using the sample covariance formula. But if there are many users who appear in week $t-1$ but not in $t$, it is unclear how to proceed. The naive approach is to estimate the  covariance using complete observations, i.e. users who appear in both weeks. However, this only works if the data is \emph{missing completely at random}. In reality, active users will show up more often and are more likely to appear in both weeks, meaning the missing data are obviously not random. 

In this section we show how the Delta method can be very useful for estimating $\cov(\bar{X}_t,\bar{X}_{t'})$ for any two time points $t$ and $t'$. We then use this to show how we can analyze within-subject studies, also known as pre-post, repeated measurement, or longitudinal studies. Our method starts with metrics directly, highly contrasting with traditional methods such as mixed effect models which build models from individual user's data. Because we study metrics directly, our model is \emph{small} and easy to solve with its complexity constant to the scale of data.\footnote{Part of the work in this section has previously appeared in a technical report \citep{guo2015flexible}. Since authoring the technical report, the authors developed a better understanding of the differences between mixed effect models and the proposed method, and we present our new findings here. The technical report has more details about other types of repeated measurement models.} 

\subsection{Methodology}

How do we estimate covariance with a completely unknown missing data mechanism? There are many existing works on handling missing data. One approach is to model the propensity of a data-point being missing using other observed predictors \citep{semiparamissing}. This requires additional covariates/predictors to be observed, plus a rather strong assumption that conditioned on these predictors, data are missing completely at random. We present a novel idea using the Delta method after \emph{data augmentation} without the need of modeling the missing data mechanism. Specifically, we use an additional indicator for the presence/absence status of a user in each period $t$. For user $i$ in period $t,$ let $I_{it} = 1$ if user $i$ appears in period $t$, and $0$ otherwise. For each user $i$ in period $t$, instead of one scalar metric value $(X_{it})$, we augment it to a vector $(I_{it}, X_{it})$. When $I_{it}=0$, i.e. user is missing, we define $X_{it}=0$. Under this simple augmentation, the metric value $\bar{X}_t$, taking the average over those non-missing measurements in period $t$, is the same as $\sum_i X_{it} / \sum_i I_{it}!$ In this connection,
\begin{equation*}
\cov(\bar{X}_t, \bar{X}_{t^\prime}) = \cov\left(\frac{\sum_i X_{it}}{\sum_i I_{it}},\frac{\sum_{i} X_{it^\prime}}{\sum_i I_{it^\prime}}\right) = \cov\left( \frac{\bar{X}_t}{\bar{I}_t}, \frac{\bar{X}_{t^\prime}}{\bar{I}_{t^\prime}} \right)
\end{equation*}
where the last equality is by dividing both numerator and denominator by the same total number of users who have appeared in any of the two periods.\footnote{Actually, if there are more than two period, we can either use only users appeared in any of the two periods, or users appeared in any of all the periods. It is mathematically the same thing if we added more users and then treat them as not appeared in $I$ and $X$, i.e. $\bar{X}_t$ remains the same.} Thanks to the central limit theorem, the vector $(\bar{I}_t,\bar{X}_t,\bar{I}_{t^\prime},\bar{X}_{t^\prime})$ is also asymptotically (multivariate) normal with covariance matrix $\Sigma$, which can be estimated using sample variance and covariance because there is \emph{no missing data} after augmentation. In Section~\ref{sec:pDelta} we already applied the Delta method to compute the variance of a ratio of metrics by taking the first order Taylor expansion. Here, we can expand the two ratios to their first order linear form 
$
(\bar{X}_t-\mu_{X_t}) / \mu_{I_t} 
- 
\mu_{X_t}(\bar{I}_t - \mu_{I_t}) / \mu_{I_t}^2,
$
where $\mu_X$ and $\mu_I$ are the means of $X$ and $I$ and the expansion for $t'$ is similar. $\cov(\bar{X}_t,\bar{X}_{t'})$ can then be computed using $\Sigma$. 

We can apply this to the general case of a within-subject study. Without loss of any generality and with the benefit of a concrete solution, we only discuss a cross-over design here. Other designs including more periods are the same in principle. In a cross-over design, we have two groups I and II. For group I, we assign them to the treatment for the first period and control for the second. For group II, the assignment order is reversed. We often pick the two groups using random selection, so each period is an A/B test by itself. 

Let $\bar{X}_{it}, i=1,2, t=1,2$ be the metric for group $i$ in period $t$. Let $\bX = (\bar{X}_{11},\bar{X}_{12},\bar{X}_{21},\bar{X}_{22})$. We know 
$
\bX \sim N (\bmu, \Sigma_\bX)
$
for a mean vector $\bmu$ and covariance matrix $\Sigma_\bX$. $\Sigma_\bX$ has the form $\text{diag}(\Sigma, \Sigma)$ since the two groups are independent with the same distribution. With the help of the Delta method, we can estimate $\Sigma$ from the data and treat it as a constant covariance matrix (hence $\Sigma_\bX$). Our model concerns the mean vector $\bmu(\btheta)$ using other parameters which represent our interest. In a cross-over design, $\btheta = (\theta_1, \theta_2, \Delta)$ where the first two are baseline means for the two periods and our main interest is the treatment effect $\Delta$. (We assume there is no treatment effect for group I carried over to the 2nd period. To study carry over effect, a more complicated design needs to be employed.) In this parameterization, 
\begin{equation}\label{eq:comodel}
\bmu(\btheta) = (\theta_1+\Delta, \theta_2, \theta_1, \theta_2+\Delta)\ .
\end{equation}

The maximum likelihood estimator and Fisher information theory \citep{VanderVaart2000} paved a general way for us to estimate $\btheta$ as well as the variance of the estimators for various mean vector models. For example, if we want to model the relative change directly, we just need to change the addition of $\Delta$ into multiplication of $(1+\Delta)$. Notice the model we are fitting here is very small, almost a toy example from a text book. All the heavy lifting is in computing $\Sigma_\bX$ which is dealt with by the Delta method where all computations are trivially distributed. When $\bmu(\btheta) = M \btheta$ is linear as in the additive cross-over model above,
$
\hat{\btheta} = (M^T\Sigma_\bX^{-1}M)^{-1}M^T\Sigma_\bX^{-1}\bX
$
and $\var(\widehat{\btheta}) =  I^{-1}$ where the Fisher Information $I = M^T\Sigma_\bX^{-1} M$.

\subsection{Numerical examples}
We simulate two groups with 1000 users each. The first group receives treatment in period 1, then control in period 2, while the second group receives the inverse. For each user, we impose an independent user effect $u_i$ that follows a normal distribution with mean 10 and standard deviation 3, and independent noises $\epsilon_{it}$ with mean 0 and standard deviation 2. Each user's base observations (before treatment effect) for the two periods are $(u_i+\epsilon_{i1}, u_i+\epsilon_{i2})$. We then model the missing data and treatment effect such that they are correlated. We define a user's engagement level $l_i$ by its user effect $u_i$ through $P(U<u_i)$, i.e. the probability that a user's $u$ is bigger than another random user. We model the treatment effect $\Delta_i$ as an additive normal with mean 10 and standard deviation 0.3, multiplied by $l_i$. For each user and each of the two periods, there is a 1 - max(0.1, $l_i$) probability of this user being missing. We can interpret the missing data as a user not showing up, or as a failure to observe. In this model, without missing data, every user has two observations and the average treatment effect should be $E(\Delta_i) = 5$ because $E(l_i) = 0.5$. Since we have missing data and it is more likely for lower engagement levels to be missing, we expect the average treatment effect for the population of all observed users to be between 5 to 10. In the current model we didn't add a time period effect such that the second period could have a different mean $\theta_2$ from the first period's mean $\theta_1$, but in analysis we always assume that this effect could exist.

We simulate the following 1000 times: each time we run both the mixed effect model $X_{it}\sim \text{IsTreatment} + \text{Time} + \text{(1|User)}$ as well as the additive cross-over model \eqref{eq:comodel} and record their estimates of the ATE and the corresponding estimated variance. We then estimate the true estimator variance using the sample variance of those estimates among 1000 trials and compare that to the mean of the estimated variance to evaluate the quality of variance estimations. We also compute the average of ATE estimates and compare to the true ATE to assess the bias. 

\begin{table}[h]
\caption{\footnotesize Simulated examples: The first three columns contain the method, true ATE and standard deviations of the corresponding methods. The last two columns contain the point estimates and average estimated standard errors.}
\centering
\resizebox{\linewidth}{!}{\begin{tabular}[t]{lrrrr}
\toprule
  & Ground Truth & SD(True) & Estimate & Avg. SE(Model)\\
\midrule
mixed effect & 6.592 & 0.1295 & 7.129 & 0.1261\\
Delta method & 6.592 & 0.1573 & 6.593 & 0.1568\\
\bottomrule
\end{tabular}}
\label{tab:crossover}
\end{table}

Table~\ref{tab:crossover} summarizes the results. We found both methods provide good variance estimation and the mixed effect model shows smaller variance. However, mixed effect also displays an upward bias in the ATE estimate while the Delta method closely tracks the true effect. To further understand the difference between the two, we separate the data into two groups: users with complete data, and users who only appear in one period (incomplete group). We run the mixed effect model for the the two groups separately. Note that in the second group each user only appears once in the data, so the model is essentially a linear model. Our working hypothesis is the following: because the mixed effect model assumes a fixed treatment effect, the effect for the complete and incomplete groups must be the same. The mixed effect model can take advantage of this assumption and construct an estimator by weighted average of the two estimates from the two groups, with the optimal weight inversely proportional to their variances. Table ~\ref{tab:co2} shows the weighted average estimator is indeed very close to mixed effect model for both estimate and variance. The weighted average is closer to the complete group because the variance there is much smaller than that of the incomplete group since within-subject comparison significantly reduces noise. This explains why the mixed effect model can produce misleading results in within-subject analysis whenever missing data patterns can be correlated with the treatment effect. The Delta method, on the other hand, offers a flexible and robust solution. 

\begin{table}[h]
\caption{\footnotesize Simulated examples: Point and variance estimates using the mixed effect vs. weighted average estimators.}
\centering
\begin{tabular}[t]{lrr}
\toprule
  & Estimate & Var\\
\midrule
mixed effect model & 7.1290 & 0.00161\\
mixed effect model on complete group & 7.3876 & 0.00180\\
linear model on incomplete group & 5.1174 & 0.01133\\
weighted avg estimator & 7.0766 & 0.00155\\
\bottomrule
\end{tabular}
\label{tab:co2}
\end{table}

\section{Concluding remarks}
\label{sec:conclusion}

\emph{Measure everything} is not only an inspiring slogan, but also a crucial step towards the holy grail of data-driven decision making. In the big data era, innovative technologies have tremendously advanced user telemetry and feedback collection, and distilling insights and knowledge from them is an imperative task for business success. To do so, we typically apply certain statistical models at the level of individual observations, fit the model using numerical procedures such as solving optimization problems, and draw conclusions and assess uncertainties from the fitted models. However, for big data such an analytical procedure can be very challenging. Based on the key observation that metrics are approximately normal due to the central limit theorem, this paper offers an alternative perspective by advocating modeling metrics, i.e., summaries or aggregations of the original data sets, in a direct manner. By doing so, we can decompose big data into small problems. However, although conceptually sound, in practice these metric level models often involve nonlinear transformations of data or complex data generating mechanisms, posing several challenges for trustworthy and efficient metric analytics.

To address these issues, we promoted the Delta method's central role in making it possible to extend the central limit theorem to new territories. We demonstrated how to apply the Delta method in four real-life applications, and illustrated how this approach naturally leads to trivially parallelizable and highly efficient implementations. Among these applications, ratio metrics, clustered randomizations and quantile metrics are all common and important scenarios for A/B testing, and business analytics in general. Within-subject studies are becoming more popular for superior sensitivities, and missing data with an unknown mechanism is ubiquitous in both the online and offline worlds. Our contribution to technical improvements, novel ideas and new understandings includes bias correction for ratio metric estimation, combination of the Delta method and outer confidence intervals for quantile metric variance estimation, and the idea of data augmentation for general missing data problems in within-subject studies. We also revealed the connection between the Delta method and mixed effect models, and explained their differences. In addition, we pointed out the advantage of the Delta method in the presence of an unknown missing data mechanism. Overall speaking, we hope this paper can serve as a practical guide of applying the Delta method in large-scale metric analyses and A/B tests, so that it is no longer just a technical detail, but a starting point and a central piece of analytical thinking.

Although the Delta method can help tackle big data problems, it does not replace the need for rigorous experimental designs and probabilistic modeling. For example, ``optimal'' choices for cluster configurations, randomization mechanisms and data transformations are known to increase the sensitivities of metrics \citep{xie2016improving, kharitonov2017learning, budylin2018consistent}. We leave them as future research directions.

\begin{acks}
We benefited from insightful discussions with Ya Xu and Daniel Ting on quantile metrics and outer confidence intervals. We thank Jong Ho Lee for implementing the efficient quantile metric estimation and confidence interval construction in ExP using Apache Spark, and Yu Guo for previous work on repeated measurements.
\end{acks}

\balance

\bibliographystyle{ACM-Reference-Format}
\bibliography{sigproc}


\begin{thebibliography}{52}


\ifx \showCODEN    \undefined \def \showCODEN     #1{\unskip}     \fi
\ifx \showDOI      \undefined \def \showDOI       #1{#1}\fi
\ifx \showISBNx    \undefined \def \showISBNx     #1{\unskip}     \fi
\ifx \showISBNxiii \undefined \def \showISBNxiii  #1{\unskip}     \fi
\ifx \showISSN     \undefined \def \showISSN      #1{\unskip}     \fi
\ifx \showLCCN     \undefined \def \showLCCN      #1{\unskip}     \fi
\ifx \shownote     \undefined \def \shownote      #1{#1}          \fi
\ifx \showarticletitle \undefined \def \showarticletitle #1{#1}   \fi
\ifx \showURL      \undefined \def \showURL       {\relax}        \fi
\providecommand\bibfield[2]{#2}
\providecommand\bibinfo[2]{#2}
\providecommand\natexlab[1]{#1}
\providecommand\showeprint[2][]{arXiv:#2}

\bibitem[\protect\citeauthoryear{Athey and Imbens}{Athey and Imbens}{2017}]%
        {athey2017econometrics}
\bibfield{author}{\bibinfo{person}{Susan Athey} {and} \bibinfo{person}{Guido~W
  Imbens}.} \bibinfo{year}{2017}\natexlab{}.
\newblock \showarticletitle{The econometrics of randomized experiments}.
\newblock \bibinfo{journal}{\emph{Handbook of Economic Field Experiments}}
  \bibinfo{volume}{1} (\bibinfo{year}{2017}), \bibinfo{pages}{73--140}.
\newblock


\bibitem[\protect\citeauthoryear{Backstrom and Kleinberg}{Backstrom and
  Kleinberg}{2011}]%
        {backstrom2011network}
\bibfield{author}{\bibinfo{person}{Lars Backstrom} {and} \bibinfo{person}{Jon
  Kleinberg}.} \bibinfo{year}{2011}\natexlab{}.
\newblock \showarticletitle{Network bucket testing}. In
  \bibinfo{booktitle}{\emph{Proceedings of the 20th international conference on
  World wide web}}. ACM, \bibinfo{pages}{615--624}.
\newblock


\bibitem[\protect\citeauthoryear{Bates, M{\"a}chler, Bolker, and Walker}{Bates
  et~al\mbox{.}}{2014a}]%
        {bates2014fitting}
\bibfield{author}{\bibinfo{person}{Douglas Bates}, \bibinfo{person}{Martin
  M{\"a}chler}, \bibinfo{person}{Ben Bolker}, {and} \bibinfo{person}{Steve
  Walker}.} \bibinfo{year}{2014}\natexlab{a}.
\newblock \showarticletitle{Fitting linear mixed-effects models using lme4}.
\newblock \bibinfo{journal}{\emph{arXiv preprint arXiv:1406.5823}}
  (\bibinfo{year}{2014}).
\newblock


\bibitem[\protect\citeauthoryear{Bates, Maechler, Bolker, Walker,
  et~al\mbox{.}}{Bates et~al\mbox{.}}{2014b}]%
        {bates2014lme4}
\bibfield{author}{\bibinfo{person}{Douglas Bates}, \bibinfo{person}{Martin
  Maechler}, \bibinfo{person}{Ben Bolker}, \bibinfo{person}{Steven Walker},
  {et~al\mbox{.}}} \bibinfo{year}{2014}\natexlab{b}.
\newblock \showarticletitle{lme4: Linear mixed-effects models using Eigen and
  S4}.
\newblock \bibinfo{journal}{\emph{R package version}} \bibinfo{volume}{1},
  \bibinfo{number}{7} (\bibinfo{year}{2014}), \bibinfo{pages}{1--23}.
\newblock


\bibitem[\protect\citeauthoryear{Boos and Hughes-Oliver}{Boos and
  Hughes-Oliver}{2000}]%
        {boos2000large}
\bibfield{author}{\bibinfo{person}{Dennis~D Boos} {and}
  \bibinfo{person}{Jacqueline~M Hughes-Oliver}.}
  \bibinfo{year}{2000}\natexlab{}.
\newblock \showarticletitle{How large does $n$ have to be for $Z$ and $t$
  intervals?}
\newblock \bibinfo{journal}{\emph{The American Statistician}}
  \bibinfo{volume}{54}, \bibinfo{number}{2} (\bibinfo{year}{2000}),
  \bibinfo{pages}{121--128}.
\newblock


\bibitem[\protect\citeauthoryear{Bottou}{Bottou}{2010}]%
        {bottou2010large}
\bibfield{author}{\bibinfo{person}{L{\'e}on Bottou}.}
  \bibinfo{year}{2010}\natexlab{}.
\newblock \showarticletitle{Large-scale machine learning with stochastic
  gradient descent}.
\newblock In \bibinfo{booktitle}{\emph{Proceedings of COMPSTAT'2010}}.
  \bibinfo{publisher}{Springer}, \bibinfo{pages}{177--186}.
\newblock


\bibitem[\protect\citeauthoryear{Brown and Wolfe}{Brown and Wolfe}{1983}]%
        {brown1983estimation}
\bibfield{author}{\bibinfo{person}{Morton~B Brown} {and}
  \bibinfo{person}{Robert~A Wolfe}.} \bibinfo{year}{1983}\natexlab{}.
\newblock \showarticletitle{Estimation of the variance of percentile
  estimates}.
\newblock \bibinfo{journal}{\emph{Computational Statistics \& Data Analysis}}
  \bibinfo{volume}{1} (\bibinfo{year}{1983}), \bibinfo{pages}{167--174}.
\newblock


\bibitem[\protect\citeauthoryear{Budylin, Drutsa, Katsev, and Tsoy}{Budylin
  et~al\mbox{.}}{2018}]%
        {budylin2018consistent}
\bibfield{author}{\bibinfo{person}{Roman Budylin}, \bibinfo{person}{Alexey
  Drutsa}, \bibinfo{person}{Ilya Katsev}, {and} \bibinfo{person}{Valeriya
  Tsoy}.} \bibinfo{year}{2018}\natexlab{}.
\newblock \showarticletitle{Consistent Transformation of Ratio Metrics for
  Efficient Online Controlled Experiments}. In
  \bibinfo{booktitle}{\emph{Proceedings of the Eleventh ACM International
  Conference on Web Search and Data Mining}}. ACM, \bibinfo{pages}{55--63}.
\newblock


\bibitem[\protect\citeauthoryear{Carpenter, Gelman, Hoffman, Lee, Goodrich,
  Betancourt, Brubaker, Guo, Li, and Riddell}{Carpenter et~al\mbox{.}}{2016}]%
        {carpenter2016stan}
\bibfield{author}{\bibinfo{person}{Bob Carpenter}, \bibinfo{person}{Andrew
  Gelman}, \bibinfo{person}{Matt Hoffman}, \bibinfo{person}{Daniel Lee},
  \bibinfo{person}{Ben Goodrich}, \bibinfo{person}{Michael Betancourt},
  \bibinfo{person}{Michael~A Brubaker}, \bibinfo{person}{Jiqiang Guo},
  \bibinfo{person}{Peter Li}, {and} \bibinfo{person}{Allen Riddell}.}
  \bibinfo{year}{2016}\natexlab{}.
\newblock \showarticletitle{Stan: A probabilistic programming language}.
\newblock \bibinfo{journal}{\emph{Journal of Statistical Software}}
  \bibinfo{volume}{20} (\bibinfo{year}{2016}), \bibinfo{pages}{1--37}.
\newblock


\bibitem[\protect\citeauthoryear{Casella and Berger}{Casella and
  Berger}{2002}]%
        {casella2002statistical}
\bibfield{author}{\bibinfo{person}{George Casella} {and}
  \bibinfo{person}{Roger~L Berger}.} \bibinfo{year}{2002}\natexlab{}.
\newblock \bibinfo{booktitle}{\emph{Statistical Inference, Second Edition}}.
\newblock \bibinfo{publisher}{Duxbury Press: Pacific Grove, CA}.
\newblock


\bibitem[\protect\citeauthoryear{Chaiken, Jenkins, Larson, Ramsey, Shakib,
  Weaver, and Zhou}{Chaiken et~al\mbox{.}}{2008}]%
        {chaiken2008scope}
\bibfield{author}{\bibinfo{person}{Ronnie Chaiken}, \bibinfo{person}{Bob
  Jenkins}, \bibinfo{person}{Per-{\AA}ke Larson}, \bibinfo{person}{Bill
  Ramsey}, \bibinfo{person}{Darren Shakib}, \bibinfo{person}{Simon Weaver},
  {and} \bibinfo{person}{Jingren Zhou}.} \bibinfo{year}{2008}\natexlab{}.
\newblock \showarticletitle{SCOPE: Easy and efficient parallel processing of
  massive data sets}.
\newblock \bibinfo{journal}{\emph{Proceedings of the VLDB Endowment}}
  \bibinfo{volume}{1} (\bibinfo{year}{2008}), \bibinfo{pages}{1265--1276}.
\newblock


\bibitem[\protect\citeauthoryear{Cortes and Vapnik}{Cortes and Vapnik}{1995}]%
        {cortes1995support}
\bibfield{author}{\bibinfo{person}{Corinna Cortes} {and}
  \bibinfo{person}{Vladimir Vapnik}.} \bibinfo{year}{1995}\natexlab{}.
\newblock \showarticletitle{Support-vector networks}.
\newblock \bibinfo{journal}{\emph{Machine Learning}}  \bibinfo{volume}{20}
  (\bibinfo{year}{1995}), \bibinfo{pages}{273--297}.
\newblock


\bibitem[\protect\citeauthoryear{Davidian, Tsiatis, and Leon}{Davidian
  et~al\mbox{.}}{2005}]%
        {semiparamissing}
\bibfield{author}{\bibinfo{person}{M. Davidian}, \bibinfo{person}{A.A.
  Tsiatis}, {and} \bibinfo{person}{S. Leon}.} \bibinfo{year}{2005}\natexlab{}.
\newblock \showarticletitle{Semiparametric Estimation of Treatment Effect in a
  Pretest-Posttest Study with Missing Data}.
\newblock \bibinfo{journal}{\emph{Statist. Sci.}}  \bibinfo{volume}{20}
  (\bibinfo{year}{2005}), \bibinfo{pages}{295--301}.
\newblock
Issue 3.


\bibitem[\protect\citeauthoryear{Deng, Lu, and Litz}{Deng
  et~al\mbox{.}}{2017}]%
        {Deng:2017}
\bibfield{author}{\bibinfo{person}{A. Deng}, \bibinfo{person}{J. Lu}, {and}
  \bibinfo{person}{J. Litz}.} \bibinfo{year}{2017}\natexlab{}.
\newblock \showarticletitle{Trustworthy analysis of online {A}/{B} tests:
  Pitfalls, challenges and solutions}. In \bibinfo{booktitle}{\emph{Proceedings
  of the Tenth ACM International Conference on Web Search and Data Mining}}.
  \bibinfo{pages}{641--649}.
\newblock


\bibitem[\protect\citeauthoryear{Deng and Shi}{Deng and Shi}{2016}]%
        {Deng:2016b}
\bibfield{author}{\bibinfo{person}{A. Deng} {and} \bibinfo{person}{X. Shi}.}
  \bibinfo{year}{2016}\natexlab{}.
\newblock \showarticletitle{Data-driven metric development for online
  controlled experiments: Seven lessons learned}. In
  \bibinfo{booktitle}{\emph{Proceedings of the 22nd ACM SIGKDD International
  Conference on Knowledge Discovery and Data Mining}}.
\newblock


\bibitem[\protect\citeauthoryear{Deng, Xu, Kohavi, and Walker}{Deng
  et~al\mbox{.}}{2013}]%
        {deng2013cuped}
\bibfield{author}{\bibinfo{person}{Alex Deng}, \bibinfo{person}{Ya Xu},
  \bibinfo{person}{Ron Kohavi}, {and} \bibinfo{person}{Toby Walker}.}
  \bibinfo{year}{2013}\natexlab{}.
\newblock \showarticletitle{Improving the sensitivity of online controlled
  experiments by utilizing pre-experiment data}. In
  \bibinfo{booktitle}{\emph{Proceedings of the 6th ACM WSDM Conference}}.
  \bibinfo{pages}{123--132}.
\newblock


\bibitem[\protect\citeauthoryear{Dmitriev, Gupta, Kim, and Vaz}{Dmitriev
  et~al\mbox{.}}{2017}]%
        {DmitrievDirtyDozen}
\bibfield{author}{\bibinfo{person}{Pavel Dmitriev}, \bibinfo{person}{Somit
  Gupta}, \bibinfo{person}{Dong~Woo Kim}, {and} \bibinfo{person}{Garnet Vaz}.}
  \bibinfo{year}{2017}\natexlab{}.
\newblock \showarticletitle{A Dirty Dozen: Twelve Common Metric Interpretation
  Pitfalls in Online Controlled Experiments}. In
  \bibinfo{booktitle}{\emph{Proceedings of the 23rd ACM SIGKDD International
  Conference on Knowledge Discovery and Data Mining}}
  \emph{(\bibinfo{series}{KDD '17})}. \bibinfo{publisher}{ACM},
  \bibinfo{address}{New York, NY, USA}, \bibinfo{pages}{1427--1436}.
\newblock
\showISBNx{978-1-4503-4887-4}
\urldef\tempurl%
\url{https://doi.org/10.1145/3097983.3098024}
\showDOI{\tempurl}


\bibitem[\protect\citeauthoryear{Dmitriev and Wu}{Dmitriev and Wu}{2016}]%
        {dmitriev2016measuring}
\bibfield{author}{\bibinfo{person}{Pavel Dmitriev} {and} \bibinfo{person}{Xian
  Wu}.} \bibinfo{year}{2016}\natexlab{}.
\newblock \showarticletitle{Measuring Metrics}. In
  \bibinfo{booktitle}{\emph{Proceedings of the 25th ACM International on
  Conference on Information and Knowledge Management}}. ACM,
  \bibinfo{pages}{429--437}.
\newblock


\bibitem[\protect\citeauthoryear{Donner}{Donner}{1987}]%
        {donner1987statistical}
\bibfield{author}{\bibinfo{person}{Allan Donner}.}
  \bibinfo{year}{1987}\natexlab{}.
\newblock \showarticletitle{Statistical methodology for paired cluster
  designs}.
\newblock \bibinfo{journal}{\emph{American Journal of Epidemiology}}
  \bibinfo{volume}{126}, \bibinfo{number}{5} (\bibinfo{year}{1987}),
  \bibinfo{pages}{972--979}.
\newblock


\bibitem[\protect\citeauthoryear{Eckles, Karrer, and Ugander}{Eckles
  et~al\mbox{.}}{2017}]%
        {eckles2017design}
\bibfield{author}{\bibinfo{person}{Dean Eckles}, \bibinfo{person}{Brian
  Karrer}, {and} \bibinfo{person}{Johan Ugander}.}
  \bibinfo{year}{2017}\natexlab{}.
\newblock \showarticletitle{Design and analysis of experiments in networks:
  Reducing bias from interference}.
\newblock \bibinfo{journal}{\emph{Journal of Causal Inference}}
  \bibinfo{volume}{5}, \bibinfo{number}{1} (\bibinfo{year}{2017}).
\newblock


\bibitem[\protect\citeauthoryear{Fan, Han, and Liu}{Fan et~al\mbox{.}}{2014}]%
        {fan2014challenges}
\bibfield{author}{\bibinfo{person}{Jianqing Fan}, \bibinfo{person}{Fang Han},
  {and} \bibinfo{person}{Han Liu}.} \bibinfo{year}{2014}\natexlab{}.
\newblock \showarticletitle{Challenges of big data analysis}.
\newblock \bibinfo{journal}{\emph{National Science Review}}
  \bibinfo{volume}{1} (\bibinfo{year}{2014}), \bibinfo{pages}{293--314}.
\newblock


\bibitem[\protect\citeauthoryear{Fieller}{Fieller}{1940}]%
        {fieller1940biological}
\bibfield{author}{\bibinfo{person}{Edgar~C Fieller}.}
  \bibinfo{year}{1940}\natexlab{}.
\newblock \showarticletitle{The biological standardization of insulin}.
\newblock \bibinfo{journal}{\emph{Supplement to the Journal of the Royal
  Statistical Society}} \bibinfo{volume}{7}, \bibinfo{number}{1}
  (\bibinfo{year}{1940}), \bibinfo{pages}{1--64}.
\newblock


\bibitem[\protect\citeauthoryear{Fieller}{Fieller}{1954}]%
        {fieller1954some}
\bibfield{author}{\bibinfo{person}{Edgar~C Fieller}.}
  \bibinfo{year}{1954}\natexlab{}.
\newblock \showarticletitle{Some problems in interval estimation}.
\newblock \bibinfo{journal}{\emph{Journal of the Royal Statistical Society.
  Series B (Methodological)}} (\bibinfo{year}{1954}),
  \bibinfo{pages}{175--185}.
\newblock


\bibitem[\protect\citeauthoryear{Fisher}{Fisher}{1922}]%
        {fisher1922mathematical}
\bibfield{author}{\bibinfo{person}{Ronald~Aylmer Fisher}.}
  \bibinfo{year}{1922}\natexlab{}.
\newblock \showarticletitle{On the mathematical foundations of theoretical
  statistics}.
\newblock \bibinfo{journal}{\emph{Philosophical Transactions of the Royal
  Society of London. Series A, Containing Papers of a Mathematical or Physical
  Character}}  \bibinfo{volume}{222} (\bibinfo{year}{1922}),
  \bibinfo{pages}{309--368}.
\newblock


\bibitem[\protect\citeauthoryear{Forero, Cano, and Giannakis}{Forero
  et~al\mbox{.}}{2010}]%
        {forero2010consensus}
\bibfield{author}{\bibinfo{person}{Pedro~A Forero}, \bibinfo{person}{Alfonso
  Cano}, {and} \bibinfo{person}{Georgios~B Giannakis}.}
  \bibinfo{year}{2010}\natexlab{}.
\newblock \showarticletitle{Consensus-based distributed support vector
  machines}.
\newblock \bibinfo{journal}{\emph{Journal of Machine Learning Research}}
  \bibinfo{volume}{11}, \bibinfo{number}{May} (\bibinfo{year}{2010}),
  \bibinfo{pages}{1663--1707}.
\newblock


\bibitem[\protect\citeauthoryear{Gelman and Hill}{Gelman and Hill}{2006}]%
        {dagelman}
\bibfield{author}{\bibinfo{person}{Andrew Gelman} {and}
  \bibinfo{person}{Jennifer Hill}.} \bibinfo{year}{2006}\natexlab{}.
\newblock \bibinfo{booktitle}{\emph{Data analysis using regression and
  multilevel/hierarchical models}}.
\newblock \bibinfo{publisher}{Cambridge University Press}.
\newblock


\bibitem[\protect\citeauthoryear{Gui, Xu, Bhasin, and Han}{Gui
  et~al\mbox{.}}{2015}]%
        {gui2015network}
\bibfield{author}{\bibinfo{person}{Huan Gui}, \bibinfo{person}{Ya Xu},
  \bibinfo{person}{Anmol Bhasin}, {and} \bibinfo{person}{Jiawei Han}.}
  \bibinfo{year}{2015}\natexlab{}.
\newblock \showarticletitle{Network {A}/{B} testing: From sampling to
  estimation}. In \bibinfo{booktitle}{\emph{Proceedings of the 24th
  International Conference on World Wide Web}}. International World Wide Web
  Conferences Steering Committee, \bibinfo{pages}{399--409}.
\newblock


\bibitem[\protect\citeauthoryear{Guo and Deng}{Guo and Deng}{2015}]%
        {guo2015flexible}
\bibfield{author}{\bibinfo{person}{Yu Guo} {and} \bibinfo{person}{Alex Deng}.}
  \bibinfo{year}{2015}\natexlab{}.
\newblock \showarticletitle{Flexible Online Repeated Measures Experiment}.
\newblock \bibinfo{journal}{\emph{arXiv preprint arXiv:1501.00450}}
  (\bibinfo{year}{2015}).
\newblock


\bibitem[\protect\citeauthoryear{Hall}{Hall}{2013}]%
        {hall2013bootstrap}
\bibfield{author}{\bibinfo{person}{Peter Hall}.}
  \bibinfo{year}{2013}\natexlab{}.
\newblock \bibinfo{booktitle}{\emph{The bootstrap and Edgeworth expansion}}.
\newblock \bibinfo{publisher}{Springer Science \& Business Media}.
\newblock


\bibitem[\protect\citeauthoryear{Hirschberg and Lye}{Hirschberg and
  Lye}{2010}]%
        {hirschberg2010geometric}
\bibfield{author}{\bibinfo{person}{Joe Hirschberg} {and} \bibinfo{person}{Jenny
  Lye}.} \bibinfo{year}{2010}\natexlab{}.
\newblock \showarticletitle{A geometric comparison of the delta and Fieller
  confidence intervals}.
\newblock \bibinfo{journal}{\emph{The American Statistician}}
  \bibinfo{volume}{64} (\bibinfo{year}{2010}), \bibinfo{pages}{234--241}.
\newblock


\bibitem[\protect\citeauthoryear{Jordan, Lee, and Yang}{Jordan
  et~al\mbox{.}}{2018}]%
        {jordan2018communication}
\bibfield{author}{\bibinfo{person}{Michael~I Jordan}, \bibinfo{person}{Jason~D
  Lee}, {and} \bibinfo{person}{Yun Yang}.} \bibinfo{year}{2018}\natexlab{}.
\newblock \showarticletitle{Communication-efficient distributed statistical
  inference}.
\newblock \bibinfo{journal}{\emph{J. Amer. Statist. Assoc.}}
  \bibinfo{volume}{in press} (\bibinfo{year}{2018}).
\newblock
\newblock
\shownote{doi:10.1080/01621459.2018.1429274.}


\bibitem[\protect\citeauthoryear{Kharitonov, Drutsa, and Serdyukov}{Kharitonov
  et~al\mbox{.}}{2017}]%
        {kharitonov2017learning}
\bibfield{author}{\bibinfo{person}{Eugene Kharitonov}, \bibinfo{person}{Alexey
  Drutsa}, {and} \bibinfo{person}{Pavel Serdyukov}.}
  \bibinfo{year}{2017}\natexlab{}.
\newblock \showarticletitle{Learning sensitive combinations of a/b test
  metrics}. In \bibinfo{booktitle}{\emph{Proceedings of the Tenth ACM
  International Conference on Web Search and Data Mining}}. ACM,
  \bibinfo{pages}{651--659}.
\newblock


\bibitem[\protect\citeauthoryear{Klar and Donner}{Klar and Donner}{2001}]%
        {klar2001current}
\bibfield{author}{\bibinfo{person}{Neil Klar} {and} \bibinfo{person}{Allan
  Donner}.} \bibinfo{year}{2001}\natexlab{}.
\newblock \showarticletitle{Current and future challenges in the design and
  analysis of cluster randomization trials}.
\newblock \bibinfo{journal}{\emph{Statistics in medicine}}
  \bibinfo{volume}{20}, \bibinfo{number}{24} (\bibinfo{year}{2001}),
  \bibinfo{pages}{3729--3740}.
\newblock


\bibitem[\protect\citeauthoryear{Kleiner, Talwalkar, Sarkar, and
  Jordan}{Kleiner et~al\mbox{.}}{2014}]%
        {kleiner2014scalable}
\bibfield{author}{\bibinfo{person}{Ariel Kleiner}, \bibinfo{person}{Ameet
  Talwalkar}, \bibinfo{person}{Purnamrita Sarkar}, {and}
  \bibinfo{person}{Michael~I Jordan}.} \bibinfo{year}{2014}\natexlab{}.
\newblock \showarticletitle{A scalable bootstrap for massive data}.
\newblock \bibinfo{journal}{\emph{Journal of the Royal Statistical Society:
  Series B (Statistical Methodology)}} \bibinfo{volume}{76},
  \bibinfo{number}{4} (\bibinfo{year}{2014}), \bibinfo{pages}{795--816}.
\newblock


\bibitem[\protect\citeauthoryear{Kohavi, Crook, Longbotham, Frasca, Henne,
  Ferres, and Melamed}{Kohavi et~al\mbox{.}}{2009a}]%
        {kohavi2009online}
\bibfield{author}{\bibinfo{person}{Ronny Kohavi}, \bibinfo{person}{Thomas
  Crook}, \bibinfo{person}{Roger Longbotham}, \bibinfo{person}{Brian Frasca},
  \bibinfo{person}{Randy Henne}, \bibinfo{person}{Juan~Lavista Ferres}, {and}
  \bibinfo{person}{Tamir Melamed}.} \bibinfo{year}{2009}\natexlab{a}.
\newblock \showarticletitle{Online experimentation at Microsoft}. In
  \bibinfo{booktitle}{\emph{Proceedings of the Third International Workshop on
  Data Mining Case Studies, held at the 5th ACM SIGKDD Conference}}.
  \bibinfo{pages}{11--23}.
\newblock


\bibitem[\protect\citeauthoryear{Kohavi, Deng, Frasca, Walker, Xu, and
  Pohlmann}{Kohavi et~al\mbox{.}}{2013}]%
        {abScale}
\bibfield{author}{\bibinfo{person}{Ron Kohavi}, \bibinfo{person}{Alex Deng},
  \bibinfo{person}{Brian Frasca}, \bibinfo{person}{Toby Walker},
  \bibinfo{person}{Ya Xu}, {and} \bibinfo{person}{Nils Pohlmann}.}
  \bibinfo{year}{2013}\natexlab{}.
\newblock \showarticletitle{Online Controlled Experiments at Large Scale}.
\newblock \bibinfo{journal}{\emph{Proceedings of the 19th ACM SIGKDD
  Conference}} (\bibinfo{year}{2013}).
\newblock


\bibitem[\protect\citeauthoryear{Kohavi, Henne, and Sommerfield}{Kohavi
  et~al\mbox{.}}{2007}]%
        {kohavi2007practical}
\bibfield{author}{\bibinfo{person}{Ron Kohavi}, \bibinfo{person}{Randal~M
  Henne}, {and} \bibinfo{person}{Dan Sommerfield}.}
  \bibinfo{year}{2007}\natexlab{}.
\newblock \showarticletitle{Practical guide to controlled experiments on the
  web: listen to your customers not to the hippo}. In
  \bibinfo{booktitle}{\emph{Proceedings of the 13th ACM SIGKDD Conference}}.
  \bibinfo{pages}{959--967}.
\newblock


\bibitem[\protect\citeauthoryear{Kohavi, Longbotham, Sommerfield, and
  Henne}{Kohavi et~al\mbox{.}}{2009b}]%
        {kohavi2009controlled}
\bibfield{author}{\bibinfo{person}{Ron Kohavi}, \bibinfo{person}{Roger
  Longbotham}, \bibinfo{person}{Dan Sommerfield}, {and}
  \bibinfo{person}{Randal~M Henne}.} \bibinfo{year}{2009}\natexlab{b}.
\newblock \showarticletitle{Controlled experiments on the web: survey and
  practical guide}.
\newblock \bibinfo{journal}{\emph{Data mining and knowledge discovery}}
  \bibinfo{volume}{18}, \bibinfo{number}{1} (\bibinfo{year}{2009}),
  \bibinfo{pages}{140--181}.
\newblock


\bibitem[\protect\citeauthoryear{Kohavi, Longbotham, and Walker}{Kohavi
  et~al\mbox{.}}{2010}]%
        {kohavi2010}
\bibfield{author}{\bibinfo{person}{R. Kohavi}, \bibinfo{person}{R. Longbotham},
  {and} \bibinfo{person}{T. Walker}.} \bibinfo{year}{2010}\natexlab{}.
\newblock \showarticletitle{Online Experiments: Practical Lessons}.
\newblock \bibinfo{journal}{\emph{Computer}} \bibinfo{volume}{43},
  \bibinfo{number}{9} (\bibinfo{date}{Sept} \bibinfo{year}{2010}),
  \bibinfo{pages}{82--85}.
\newblock


\bibitem[\protect\citeauthoryear{Krewski}{Krewski}{1976}]%
        {krewski1976distribution}
\bibfield{author}{\bibinfo{person}{Daniel Krewski}.}
  \bibinfo{year}{1976}\natexlab{}.
\newblock \showarticletitle{Distribution-free confidence intervals for quantile
  intervals}.
\newblock \bibinfo{journal}{\emph{J. Amer. Statist. Assoc.}}
  \bibinfo{volume}{71}, \bibinfo{number}{354} (\bibinfo{year}{1976}),
  \bibinfo{pages}{420--422}.
\newblock


\bibitem[\protect\citeauthoryear{Liang and Zeger}{Liang and Zeger}{1986}]%
        {liang1986longitudinal}
\bibfield{author}{\bibinfo{person}{Kung-Yee Liang} {and}
  \bibinfo{person}{Scott~L Zeger}.} \bibinfo{year}{1986}\natexlab{}.
\newblock \showarticletitle{Longitudinal data analysis using generalized linear
  models}.
\newblock \bibinfo{journal}{\emph{Biometrika}} \bibinfo{volume}{73},
  \bibinfo{number}{1} (\bibinfo{year}{1986}), \bibinfo{pages}{13--22}.
\newblock


\bibitem[\protect\citeauthoryear{Meyer}{Meyer}{1987}]%
        {meyer1987outer}
\bibfield{author}{\bibinfo{person}{John~S Meyer}.}
  \bibinfo{year}{1987}\natexlab{}.
\newblock \showarticletitle{Outer and inner confidence intervals for finite
  population quantile intervals}.
\newblock \bibinfo{journal}{\emph{J. Amer. Statist. Assoc.}}
  \bibinfo{volume}{82}, \bibinfo{number}{397} (\bibinfo{year}{1987}),
  \bibinfo{pages}{201--204}.
\newblock


\bibitem[\protect\citeauthoryear{Rudin et~al\mbox{.}}{Rudin
  et~al\mbox{.}}{1964}]%
        {rudin1964principles}
\bibfield{author}{\bibinfo{person}{Walter Rudin} {et~al\mbox{.}}}
  \bibinfo{year}{1964}\natexlab{}.
\newblock \bibinfo{booktitle}{\emph{Principles of mathematical analysis}}.
  Vol.~\bibinfo{volume}{3}.
\newblock \bibinfo{publisher}{McGraw-hill New York}.
\newblock


\bibitem[\protect\citeauthoryear{Tang, Agarwal, O'Brien, and Meyer}{Tang
  et~al\mbox{.}}{2010}]%
        {googlesurvey}
\bibfield{author}{\bibinfo{person}{Diane Tang}, \bibinfo{person}{Ashish
  Agarwal}, \bibinfo{person}{Deirdre O'Brien}, {and} \bibinfo{person}{Mike
  Meyer}.} \bibinfo{year}{2010}\natexlab{}.
\newblock \showarticletitle{Overlapping Experiment Infrastructure: More,
  Better, Faster Experimentation}.
\newblock \bibinfo{journal}{\emph{Proceedings of the 16th ACM SIGKDD
  Conference}} (\bibinfo{year}{2010}).
\newblock


\bibitem[\protect\citeauthoryear{{Van der Vaart}}{{Van der Vaart}}{2000}]%
        {VanderVaart2000}
\bibfield{author}{\bibinfo{person}{Aad~W {Van der Vaart}}.}
  \bibinfo{year}{2000}\natexlab{}.
\newblock \bibinfo{booktitle}{\emph{{Asymptotic statistics}}}.
  Vol.~\bibinfo{volume}{3}.
\newblock \bibinfo{publisher}{Cambridge university press}.
\newblock
\showISBNx{0521784506}


\bibitem[\protect\citeauthoryear{Von~Luxburg and Franz}{Von~Luxburg and
  Franz}{2009}]%
        {von2009geometric}
\bibfield{author}{\bibinfo{person}{Ulrike Von~Luxburg} {and}
  \bibinfo{person}{Volker~H Franz}.} \bibinfo{year}{2009}\natexlab{}.
\newblock \showarticletitle{A geometric approach to confidence sets for ratios:
  Fieller's theorem, generalizations and bootstrap}.
\newblock \bibinfo{journal}{\emph{Statistica Sinica}} (\bibinfo{year}{2009}),
  \bibinfo{pages}{1095--1117}.
\newblock


\bibitem[\protect\citeauthoryear{Wang and Zhou}{Wang and Zhou}{2012}]%
        {wang2012distributed}
\bibfield{author}{\bibinfo{person}{Dongli Wang} {and} \bibinfo{person}{Yan
  Zhou}.} \bibinfo{year}{2012}\natexlab{}.
\newblock \showarticletitle{Distributed support vector machines: An overview}.
  In \bibinfo{booktitle}{\emph{Control and Decision Conference (CCDC), 2012
  24th Chinese}}. IEEE, \bibinfo{pages}{3897--3901}.
\newblock


\bibitem[\protect\citeauthoryear{Wasserman}{Wasserman}{2003}]%
        {allofstat}
\bibfield{author}{\bibinfo{person}{Larry Wasserman}.}
  \bibinfo{year}{2003}\natexlab{}.
\newblock \bibinfo{booktitle}{\emph{All of Statistics: A Concise Course in
  Statistical Inference}}.
\newblock \bibinfo{publisher}{Springer}.
\newblock


\bibitem[\protect\citeauthoryear{Xie and Aurisset}{Xie and Aurisset}{2016}]%
        {xie2016improving}
\bibfield{author}{\bibinfo{person}{Huizhi Xie} {and} \bibinfo{person}{Juliette
  Aurisset}.} \bibinfo{year}{2016}\natexlab{}.
\newblock \showarticletitle{Improving the sensitivity of online controlled
  experiments: Case studies at netflix}. In
  \bibinfo{booktitle}{\emph{Proceedings of the 22nd ACM SIGKDD International
  Conference on Knowledge Discovery and Data Mining}}. ACM,
  \bibinfo{pages}{645--654}.
\newblock


\bibitem[\protect\citeauthoryear{Xu, Chen, Fernandez, Sinno, and Bhasin}{Xu
  et~al\mbox{.}}{2015}]%
        {xu2015infrastructure}
\bibfield{author}{\bibinfo{person}{Ya Xu}, \bibinfo{person}{Nanyu Chen},
  \bibinfo{person}{Addrian Fernandez}, \bibinfo{person}{Omar Sinno}, {and}
  \bibinfo{person}{Anmol Bhasin}.} \bibinfo{year}{2015}\natexlab{}.
\newblock \showarticletitle{From infrastructure to culture: A/{B} testing
  challenges in large scale social networks}. In
  \bibinfo{booktitle}{\emph{Proceedings of the 21th ACM SIGKDD International
  Conference on Knowledge Discovery and Data Mining}}. ACM,
  \bibinfo{pages}{2227--2236}.
\newblock


\bibitem[\protect\citeauthoryear{Zaharia, Xin, Wendell, Das, Armbrust, Dave,
  Meng, Rosen, Venkataraman, Franklin, et~al\mbox{.}}{Zaharia
  et~al\mbox{.}}{2016}]%
        {zaharia2016apache}
\bibfield{author}{\bibinfo{person}{Matei Zaharia}, \bibinfo{person}{Reynold~S
  Xin}, \bibinfo{person}{Patrick Wendell}, \bibinfo{person}{Tathagata Das},
  \bibinfo{person}{Michael Armbrust}, \bibinfo{person}{Ankur Dave},
  \bibinfo{person}{Xiangrui Meng}, \bibinfo{person}{Josh Rosen},
  \bibinfo{person}{Shivaram Venkataraman}, \bibinfo{person}{Michael~J
  Franklin}, {et~al\mbox{.}}} \bibinfo{year}{2016}\natexlab{}.
\newblock \showarticletitle{Apache Spark: A unified engine for big data
  processing}.
\newblock \bibinfo{journal}{\emph{Commun. ACM}} \bibinfo{volume}{59},
  \bibinfo{number}{11} (\bibinfo{year}{2016}), \bibinfo{pages}{56--65}.
\newblock


\bibitem[\protect\citeauthoryear{Zinkevich, Weimer, Li, and Smola}{Zinkevich
  et~al\mbox{.}}{2010}]%
        {zinkevich2010parallelized}
\bibfield{author}{\bibinfo{person}{Martin Zinkevich}, \bibinfo{person}{Markus
  Weimer}, \bibinfo{person}{Lihong Li}, {and} \bibinfo{person}{Alex~J Smola}.}
  \bibinfo{year}{2010}\natexlab{}.
\newblock \showarticletitle{Parallelized stochastic gradient descent}. In
  \bibinfo{booktitle}{\emph{Advances in Neural Information Processing
  Systems}}. \bibinfo{pages}{2595--2603}.
\newblock


\end{thebibliography}

\end{document}